\documentclass[twoside,twocolumn,aps,pra]{revtex4-1}
\usepackage[T1]{fontenc}
\usepackage[latin9]{inputenc}
\usepackage{amsmath}
\usepackage{amssymb}
\usepackage{graphicx}

\makeatletter

\@ifundefined{textcolor}{}
{%
 \definecolor{BLACK}{gray}{0}
 \definecolor{WHITE}{gray}{1}
 \definecolor{RED}{rgb}{1,0,0}
 \definecolor{GREEN}{rgb}{0,1,0}
 \definecolor{BLUE}{rgb}{0,0,1}
 \definecolor{CYAN}{cmyk}{1,0,0,0}
 \definecolor{MAGENTA}{cmyk}{0,1,0,0}
 \definecolor{YELLOW}{cmyk}{0,0,1,0}
}

\usepackage{bm}\usepackage{amsthm}
\usepackage{epstopdf}\usepackage{txfonts}

\newcommand{\1}{\leavevmode{\rm 1\ifmmode\mkern  -4.8mu\else\kern -.3em\fi I}}

\DeclareRobustCommand{\openzero}{\leavevmode\hbox{0\kern-.55em0}}
\mathchardef\minus="002D

\makeatother

\begin{document}

\title{Geometry, robustness, and emerging unitarity in dissipation-projected dynamics}

\author{Paolo Zanardi}
\author{Lorenzo Campos Venuti}
\affiliation{Department of Physics and Astronomy, and Center for Quantum Information
Science \& Technology, University of Southern California, Los Angeles,
CA 90089-0484}
\begin{abstract}
Quantum information can be encoded in the set of steady-states (SSS)
of a driven-dissipative system. Non steady-states are separated by a
large dissipative gap that adiabatically decouples them way while
the dynamics inside the SSS is governed by an effective, dissipation-projected,
Hamiltonian. The latter results from a highly non-trivial interplay
between a weak driving with the fast relaxation process that continuously
projects the system back to the SSS. This amounts to a novel type
of environment-induced quantum Zeno effect. We prove that the dissipation-projected
dynamics is of geometric nature and that it is robust against different
types of hamiltonian and dissipative perturbations. Remarkably, in
some cases an effective unitary dynamics can emerge out of purely
dissipative interactions. 
\end{abstract}
\maketitle

\section{Introduction}

Since the earliest days of quantum
information processing (QIP) weak coupling to the environmental degrees
of freedom has been regarded as one of the essential prerequisites.
In fact decoherence and dissipation generally spoil the unitary character
of the quantum dynamics and induce errors into the computational process.
In order to overcome such an obstacle a variety of techniques have
been devised including quantum error correction \cite{QEC}, decoherence-free
subspaces (DFSs) \cite{DFS,DFS1,DFS-exp}, noiseless subsystems
(NS) \cite{NS,stab,NS-top,NS-exp} and geometric/holonomic quantum computation \cite{Jones, HQC, HQC-science}.

However, it has been recently realized that dissipation and decoherence
may even play a positive role to the aim of coherent quantum manipulations.
Indeed, it has been shown that, properly engineered, dissipative dynamics
can in principle be used to enact QIP primitives such as quantum state
preparation \cite{Kraus-prep,kastoryano2011dissipative}, quantum
simulation \cite{barreiro2011open} and computation \cite{verstraete2009quantum}.
Exotic physical properties such as topological order \cite{dissi-top}
and non-abelian synthetic gauge fields \cite{gauge} can also be achieved
by engineered dissipation.

In a nutshell the idea is that one can design {\em{driven-dissipative}}
systems such that their {\em{steady-states}} enjoy some computationally
desirable property. For example in Ref.~\cite{kastoryano2011dissipative}
the unique steady-state is maximally entangled, while in Ref.~\cite{verstraete2009quantum}
the steady states encode for an arbitrary quantum computation! Moreover,
the irreversible and attractive nature of dissipative dynamics endows
these techniques with a degree of robustness against imperfections
in preparation and control. All this leads to a dramatic paradigm
shift in QIP: {\em{ noise and dissipation should not be viewed
as detrimental but may in fact be considered as a resource}}. 

In this paper we will build upon our recent discovery on how to enact
coherent dynamics over the set of steady states (SSS) of a strongly
dissipative system \cite{zanardi-dissipation-2014}. Quantum information
is encoded in sectors of the SSS while non steady-states are separated
by the large dissipative gap that adiabatically decouples them away.
A weak Hamiltonian control gives rise to an effective dynamics {\em{inside}}
the SSS that is ruled by a dissipation-projected Hamiltonian. The
latter results from a highly non-trivial interplay between the control
and the fast relaxation process that continuously projects the system
back onto the SSS. This amounts to a novel type of environment-induced
quantum Zeno effect \cite{zeno-subspaces,dan-zeno}. 

In this paper we will show  that the dissipation-projected dynamics is geometric in nature.
This means that this approach can be regarded as a dissipative extension
of the fault-tolerant techniques of geometric and holonomic quantum
computation \cite{Jones,HQC,HQC-science}. We will also prove that the
dissipation-projected Hamiltonians are protected against several types
of perturbations (unitary and dissipative) and may allow for  robust QIP.
Finally we will show how an effective unitary evolution may emerge out of  suitable
dissipative perturbations of a purely dissipative dynamics.
This ``emerging unitarity" phenomenon is perhaps the single most surprising one of our results.



\section{The dissipation-projection theorem}


We will consider quantum open systems whose dynamics is described
by the equation 
\begin{equation}
\frac{d\rho(t)}{dt}={\cal L}\,\rho(t).\label{basic-eq}
\end{equation}
The superoperator ${\cal L}$ will be referred as to the Liouvillian
An open quantum system generically admits a unique {\em{steady-state}}
$\rho_{\infty}$ that is approached by the time-evolving density matrix
$\rho(t)$ as the time goes to infinity. Asymptotically the information-theoretic
distance $D(\rho(t),\,\rho_{\infty}):=\frac{1}{2}\|\rho(t)-\rho_{\infty}\|_{1}$
decays exponentially with time 
where the time-scale $\tau_{R}$
is referred to as the {\em{relaxation time}}. For $t\gg\tau_{R}$
the time-evolved state becomes indistinguishable from the steady
state. According to Eq.~(\ref{basic-eq}) the steady state satisfies
${\cal L}(\rho_{\infty})=0$ i.e.~it lies lies in the kernel
of the Liouvillian. Uniqueness of the steady-state translates into
a one-dimensional kernel. In this paper we will focus on the case
in which the Liouvillian can be decomposed as ${\cal L}={\cal L}_{0}+{\cal L}_{1}$
in such a way that 
\begin{itemize}
\item {\bf{i)}} The relaxation time of ${\cal L}_{0}$ is the shortest
time-scale of the problem. Equivalently, the dissipative gap of ${\cal L}_{0}$, 
$\tau_{R}^{-1}$,  is the largest energy scale.
\item {\bf{ii)}} The kernel of ${\cal L}_{0}$ is {\em{high-dimensional}}
and attractive (the non-zero eigenvalues of ${\cal L}_{0}$ have a
negative real part) 
\end{itemize}
We will denote by ${\cal P}_{0}$ (${\cal Q}_{0}=1-{\cal P}_{0}$) the
projection onto the kernel of ${\cal L}_{0}$ (its complementary).
The steady-state set (SSS) is given by those states $\rho$ such
that ${\cal P}_{0}(\rho)=\rho.$ The critical assumption is that the
SSS is high-dimensional. A prototypical instance of this {\em{non-generic}}
situation is the following:
\vskip 0.2truecm

{\em{Example 0.-}} Suppose a system $S$ is joined to a system $B$
and that the dissipation acts only on the latter. Let $\rho_{B}$
denote the (generically) unique steady-state of $B$ and by $\rho$
any state of $S$. It is then obvious that any bi-partite state of
the form $\rho\otimes\rho_{B}$ is a steady-state of the full dynamical
system when the $S$ and $B$ are decoupled. Clearly, any transformation
over $S$ is a symmetry of the dynamics. For the sake of concreteness
one may think of a two-level atom $S$ strongly coupled to a leaky
cavity mode $B$. To a good approximation dissipation acts directly
just on $B.$ Formally, the Hilbert space is ${\cal H}={\cal H}_{S}\otimes{\cal H}_{B}$
and ${\cal L}_{0}={\mathbf{1}}_{S}\otimes{\cal L}_{B}$ where the
Liouvillian ${\cal L}_{B}$ admits a {\em{unique}} steady-state
$\rho_{B}$. In this case ${\cal P}_{0}(X)={\mathrm{Tr}}_{B}(X)\otimes\rho_{B},$
the kernel of ${\cal L}_{0}$ has dimension $({\mathrm{dim}}\,{\cal H}_{S})^{2},$
and the SSS can be identified with the state-space of $S.$ This apparently
trivial example will be later considerably generalized resorting to
the theory of NSs \cite{NS}. 

The fundamental technical result we would like to build upon is the
following fact proved in \cite{zanardi-dissipation-2014} (see also Sec.~A): 
\vskip 0.2truecm
\textbf{Projection Theorem.-} 
Suppose ${\cal L}={\cal L}_0+ {\cal L}_1$ with $\|{\cal L}_1\|=O(1/T)$ then 
\begin{equation}
\sup_{t\in[0,T]}\|({\cal E}_{t}-e^{t {{\cal L}}_{{\textrm{\textrm{eff}}}}}){\cal P}_{0}\|=O(1/{T})\label{projection-th}
\end{equation}
where ${{\cal L}}_{ {\textrm{eff}}}:={\cal P}_{0}\,{{\cal L}}\,{\cal P}_{0}= {\cal P}_{0}\,{{\cal L}_1}\,{\cal P}_{0}$
and ${\cal E}_{t}=e^{t {\cal L}}.$ 
 \vskip 0.2truecm

\noindent In words: if the system is prepared at time $t=0$ inside
the SSS then, in the large $T$ limit, the time-evolution leaves the
SSS invariant and it is governed by the effective generator ${\cal L}_{{\textrm{eff}}}$.

In several of the applications we will discuss below the perturbation
will be of Hamiltonian type i.e., ${\cal L}_{1}=-i[K,\bullet],\,(K=K^{\dagger});$
in that case it will be denoted by ${\cal K}.$ 
The key point is that ${\cal K}_{{\textrm{eff}}}={\cal P}_{0}\,{{\cal K}}\,{\cal P}_{0}$ turns out to be
an Hamiltonian; it will be referred to as the {\em{dissipation-projected}}
Hamiltonian. Physically, this means that strong dissipation, while
dressing the Hamiltonian by a continuous projection onto the SSS,
does not alter its unitary character. Non steady-states are adiabatically
decoupled away. The SSS and unitarity are protected by the large dissipative
gap of ${\cal L}_{0}.$

For example, in the  {\em{Example 0}}  discussed above, where
the Liouvillian ${\cal L}_{B}$ has a {\em{unique}} steady-state
$\rho_{B}$, one finds ${\cal K}_{{\textrm{eff}}}=-i[K_{{\textrm{eff}}},\,\bullet]$
where $K_{{\textrm{eff}}}={\mathrm{Tr}}_{B}(K\rho_{B})\otimes{\mathbf{1}}_{B}$.
We see that in fact ${\cal K}_{{\textrm{eff}}}$ is Hamiltonian.



In Sec.~\ref{sec:Error-estimate} we prove Eq.~(\ref{projection-th})
and we give a rigorous estimate for the coefficient in its RHS. It
turns out {[}see Eq.~(\ref{eq:bound_final-1}){]} that the numerical
factor is $c\tau_{R}$ where $\tau_{R}$ is the relaxation time of
the unperturbed dynamics and $c$ is a $O(1)$ constant. This fact
is important as it implies that the error can be made small, either
by making $T$ larger (which also makes the waiting time $O(T)$ longer)
or by making dissipation faster (i.e.~$\tau_{R}$ smaller). Indeed,
measuring times in unit of $\tau_{R}$ one realizes that the expansion
parameter in Eq.~(\ref{projection-th}) is really $\tau_{R}/T$.
In other terms the ``long $T$ limit'' just means that the {Hamiltonian
norm has to be much smaller than the dissipative gap} {[}$=O(\tau_{R}^{-1})]$.
The latter represents the physical quantity that in real applications
has to be engineered in order to make it as large as possible. Equivalently,
one wants to make the relaxation time $\tau_{R}$ as short as possible.
We have to operate in the deep dissipative regime.

%
\section{Dissipative Holonomies}

Let us now discuss the intimate relation between our basic result (\ref{projection-th})
and geometric and holonomic quantum computation \cite{Jones,HQC}.
We will show that the effective evolution (\ref{projection-th}) is
in fact {\em{geometric}} and is given by a super-operator holonomy.

The possibility of merging dissipation dynamics and holonomic quantum
computation \cite{HQC,HQC1,HQC-science} by reservoir engineering
was first suggested in Refs.~\cite{angelo,ogy}. More specifically,
in \cite{angelo} a time-dependent Lindbladian dynamics admitting
a DFS was considered, and it was shown that under a suitable adiabatic
condition, a state initially in a DFS remains inside the subspace
and, hence, is rigidly transported around the Hilbert space together
with the DFS. The evolution is, in fact, coherent, although entirely
produced by an incoherent phenomenon. Moreover, when the DFS eventually
returns to its initial configuration, the net effect is a holonomic
transformation on the states in the subspace. Counterintuitively,
the effect of the dissipation on the (time-dependent) DFS can be made
smaller by making the dissipation rate larger. The authors qualitatively
explain this phenomenon in terms of some sort of environment-induced
quantum Zeno effect where the action of a strong environment can be
regarded as a measuring apparatus continuously monitoring the slowly
moving DFS.

In order to establish a connection between these findings and the
results we have discussed so far it suffices to move to a rotated
reference frame by defining $\tilde{\rho}(t):={\cal U}_{t}^{\dagger}\rho(t)$
where ${\cal U}_{t}(X):=e^{t{\cal K}}(X)=e^{-itK}Xe^{itK}.$ {{
In this rotated frame $\tilde{\rho}(t)$ evolves in a {\em{time-dependent}}
bath 
\begin{equation}
\frac{d\tilde{\rho}(t)}{dt}={\cal L}_{t}\,\tilde{\rho}(t),\qquad{\cal L}_{t}:={\cal U}_{t}^{\dagger}{\cal L}_{0}{\cal U}_{t}\label{rot-eq}
\end{equation}
In the rotated frame the dynamical semi-group is given by $\tilde{{\cal E}}_{t}={\cal U}_{t}^{\dagger}{\cal E}_{t}$
and a state $\tilde{\rho}_{t}$ is an {\em{instantaneous}} steady-state
of ${\cal L}_{t}$ iff $\tilde{\rho}_{t}={\cal U}_{t}^{\dagger}\rho_{0}$
where $\rho_{0}$ is a steady-state of ${\cal L}_{0}.$ It follows
that the projector onto the kernel of ${\cal L}_{t}$ is given by
${\cal P}_{t}={\cal U}_{t}^{\dagger}{\cal P}_{0}{\cal U}_{t}=e^{-t{\cal K}}{\cal P}_{0}e^{t{\cal K}}.$
Moreover, in the rotated-frame the dissipation-projected dynamics
is {\em{geometric}}.
\vskip 0.2truecm
{\bf{Proposition 1.--}} a) The Projection Theorem (\ref{projection-th})
can be reformulated in the form 
\begin{equation}
\|\tilde{{\cal E}}_{t}\,{\cal P}_{0}-{\mathbf{T}}\exp(\int_{0}^{t}d\tau\,[\dot{{\cal P}}_{\tau},\,{\cal P}_{\tau}])\,{\cal P}_{0}\|=O(1/T)
\label{HQC-rot}
\end{equation}
where ${\mathbf{T}}$ denotes the chronological ordering symbol. 

b) The ${\mathbf{T}}$-ordered
geometric superoperator in (\ref{HQC-rot}) can be rewritten as
\begin{equation}
X(t)=\lim_{N\to\infty}{\mathbf{T}}\prod_{j=1}^{N}{\cal P}_{t_{j}}=e^{-t{\cal K}}\lim_{N\to\infty}\left(e^{\frac{t}{N}{\cal K}}{\cal P}_{0}\right)^{N}\label{projection-string}
\end{equation}
where $t_{j}=j{t}/{N},\, j=0,\ldots,N$. Namely
the evolution corresponds to an infinite, time-ordered, succession
of projections onto the instantaneous SSS. Equivalently, to a succession
of ${\cal P}_{0}$ {\em{interleaved}} with infinitesimal unitaries
evolutions $e^{\frac{1}{N}{\cal K}}.$

{\em{Proof.--}} a) From unitarity of ${\cal U}_{t}=e^{t{\cal K}}$
and Eq. ~(\ref{projection-th}) one has 
\begin{equation}
\|{\cal E}_{t}{\cal P}_{0}-e^{t{\cal P}_{0}{\cal K}{\cal P}_{0}}{\cal P}_{0}\|=\|\tilde{{\cal E}}_{t}{\cal P}_{0}-X(t)\|=O(1/T)\label{norm-rot-frame}
\end{equation}
where $X(t):=e^{-t{\cal K}}e^{t{\cal P}_{0}{\cal K}{\cal P}_{0}}{\cal P}_{0}.$
By differentiation 
\begin{multline}
\dot{X}(t)=-{\cal K}e^{-t{\cal K}}e^{t{\cal P}_{0}{\cal K}{\cal P}_{0}}{\cal P}_{0}+e^{-t{\cal K}}{\cal P}_{0}{\cal K}{\cal P}_{0}e^{t{\cal P}_{0}{\cal K}{\cal P}_{0}}{\cal P}_{0}=\\
-{\cal K}{\cal P}_{t}X(t)+{\cal P}_{t}{\cal K}X(t)=[-{\cal K},\,{\cal P}_{t}]X(t)=\dot{{\cal P}}_{t}\, X(t)
\end{multline}
Notice also that ${\cal P}_{t}X(t)=X(t)$ and ${\cal P}_{t}\dot{{\cal P}}_{t}{\cal P}_{t}=0,$
whence $\dot{X}(t)=\dot{{\cal P}}_{t}\, X(t)=\dot{{\cal P}}_{t}{\cal P}_{t}\, X(t)=(\dot{{\cal P}}_{t}{\cal P}_{t}-{\cal P}_{t}\dot{{\cal P}}_{t})\,{\cal P}_{t}X(t),$
namely 
\begin{equation}
\frac{dX(t)}{dt}=[\dot{{\cal P}}_{t},\,{\cal P}_{t}]\, X(t)\Rightarrow X(t)={\mathbf{T}}\, e^{\int_{0}^{t}d\tau\,[\dot{{\cal P}}_{\tau},\,{\cal P}_{\tau}]}X(0).
\end{equation}
Eq.~(\ref{HQC-rot}) is now obtained by using (\ref{norm-rot-frame})
and $X(0)={\cal P}_{0}.$ 

b)  Proceeding formally, if $\tilde{X}(t)=\prod_{\tau\in[0,\, t]}{\cal P}_{\tau}$
then 
\begin{multline}
\tilde{X}(t+dt)-\tilde{X}(t)={\cal P}_{t+dt}\tilde{X}(t)-\tilde{X}(t)=\\
({\cal P}_{t+dt}-{\cal P}_{t})\tilde{X}(t)=\dot{{\cal P}}_{t}\tilde{X}(t)dt+O(dt^{2}).
\end{multline}
Whence $\dot{\tilde{X}}(t)=\dot{{\cal P}}_{t}\tilde{X}(t)$. Since
$X(t)$ and $\tilde{X}(t)$ fulfill the same ODE and the same initial
condition $X(0)=\tilde{X}(0)={\cal P}_{0}$ they have to be the same
function. This proves the first equality in (\ref{projection-string})
while the second can be verified by direct inspection using the definition
of the ${\cal P}_{t}$'s. $\hfill\Box$ 
\vskip 0.2truecm
The integral
in (\ref{HQC-rot}) is clearly invariant under time reparametrizations
$\tau\to\tau^{\prime}=\tau^{\prime}(\tau)$ and it is therefore of
{{geometric}} nature i.e., it depends only on the path $t\to{\cal P}_{t}$
in the space of (super) projections. We also see that the super-operator
holonomy is the line integral of the ``tautological'' connection
${\cal A}=[\dot{{\cal P}}(\tau),\,{\cal P}(\tau)]$ \cite{Nak}.


If one replaces in Eq.~(\ref{projection-string}) the projection
${\cal P}_{0}$ with a more-general CP map e.g., generalized measurement,
basically {\em{all}} the Quantum-Zeno like QIP protocols recently
discussed in the literature are recovered \cite{ogy,HQC-zeno,cave,kwek-zeno}.
In all these works the geometric and holonomic nature of the resulting
dynamics have been discussed on the basis of the particular case at
hand, and a general comprehensive theoretical understanding seems
to be lacking. The dissipation-projection theorem (\ref{projection-th})
seems able to provide such an underlying conceptual framework.


\section{SSS and Interaction Algebras }

In this section we would like to discuss an important class of dissipative
systems whose SSS can be fully characterized on general algebraic
grounds and at the same time describes physically relevant cases.

Let us consider the most general dissipative generator ${\cal L}_{0}$
of a Markovian quantum dynamical semi-group ${\cal E}_{t}:=e^{t\,{\cal L}_{0}}$.
Thanks to the Lindblad theorem \cite{Lindblad-paper} the Liouvillian
can be written as 
\begin{equation}
{\cal L}_{0}(\rho)=\sum_{\alpha}(L_{\alpha}\rho L_{\alpha}^{\dagger}-\frac{1}{2}\{L_{\alpha}^{\dagger}L_{\alpha},\,\rho\})
\label{Lindblad-eq}
\end{equation}
The $L_{\alpha}$'s are the co-called Lindblad operators. Let us now
define two operator algebras associated with (\ref{Lindblad-eq})
\begin{equation}
{\cal A}:={\mathrm{Alg}}[\{L_{\alpha},\, L_{\alpha}^{\dagger}\}_{\alpha}],\qquad{\cal A}^{\prime}:=\{X\in{\mathbf{L}}({\cal H})\,/\,[X,\,{\cal A}]=0\}\label{algs}
\end{equation}
The algebra ${\cal A}$ is the associative algebra generated by the
Lindblad operators $L_{\alpha}$ and their hermitian conjugates, it
will be referred to as the {\em{interaction algebra}} \cite{NS}
and ${\cal A}^{\prime}$ as its {\em{commutant}}. These algebras
play a fundamental role in unifying all the quantum information stabilization
techniques developed so far \cite{stab,NS-top}. Both ${\cal A}$
and ${\cal A}^{\prime}$ are closed under hermitian conjugation and
can be regarded as (finite-dimensional) $C^{*}-$algebras. Standard
structure-theorems then imply that the state-space breaks down into
$d_{J}$--dimensional irreducible representations of ${\cal A}$ (labeled
by $J$) each of them appearing with multiplicity $n_{J}$: 
\begin{equation}
{\cal H}\cong\bigoplus_{J}{\mathbf{C}}^{n_{J}}\otimes{\mathbf{C}}^{d_{J}}.\label{state-space-decomp}
\end{equation}
From this it follows that at the algebra level one has 
\begin{equation}
{\cal A}\cong\bigoplus_{J}\openone_{n_{J}}\otimes{\mathrm{L}}({\mathbf{C}}^{d_{J}}),\qquad{\cal A}^{\prime}\cong\bigoplus_{J}{\mathrm{L}}({\mathbf{C}}^{n_{J}})\otimes\openone_{d_{J}}.\label{alg-decomp}
\end{equation}
From the first Eq.~in (\ref{alg-decomp})
it follows that the Liouvillian (\ref{Lindblad-eq}) preserves the
direct-sum structure of the Hilbert space i.e., ${\cal L}$ is $J$
block-diagonal, and that it has a trivial action on the ${\mathbf{C}}^{n_{J}}$
factors. For this reason the latter are termed ``noiseless-subsystems''
and is where quantum information can be stored safely from the influence
of the environment described by of ${\cal L}_{0}$ (more on this in
the next section) \cite{NS}. In each of the ${\cal L}_{0}$-invariant
$J$ blocks the situation coincides with the one of 
{\em{Example 0}}. In other terms Eq.~(\ref{Lindblad-eq}) corresponds to a direct
sum of bi-partite systems in which the noise acts just on one of the
two (virtual) subsystems i.e., ${\mathbf{C}}^{d_{J}}.$ In this sense
this class of models can be regarded as a far reaching generalization
of {\em{Example 0}}   \cite{zanardi-dissipation-2014}.

We now assume that $\sum_{\alpha}[L_{\alpha},\, L_{\alpha}^{\dagger}]=0.$
Under these assumptions the dynamical semi-group $\{e^{t{\cal L}_{0}}\}_{t\ge0}$
leaves the identity fixed as ${\cal L}_{0}(\openone)=0$ and Ker$\,{\cal L}_{0}={\cal A}^{\prime}$
\cite{Kribs}. Such a ${\cal L}_{0}$ will be referred to as {\em{unital}}.
From the second Eq. in (\ref{alg-decomp}) we see that the SSS is
given by the convex hull of states of the form $\omega_{J}\otimes\openone_{d_{J}}/d_{J}$
where $\omega_{J}$ is a state over the factor ${\mathbf{C}}^{n_{J}}$.
Since Ker$\,{\cal L}_{0}={\cal A}^{\prime}$ it follows that ${\cal P}_{0}$
is the projection onto the commutant algebra ${\cal A}^{\prime}$,
namely \cite{zanardi-dissipation-2014} 
\begin{equation}
{\cal P}_{0}(X)=\int dU\, UXU^{\dagger}=\sum_{J}{\mathrm{Tr}}_{d_{J}}\left(\Pi_{J}\, X\,\Pi_{J}\right)\otimes\openone_{d_{J}}/d_{J}\in{\cal A}^{\prime}\label{alg-projection}
\end{equation}
where the Haar-measure integral is performed over the unitary group of the algebra ${\cal A}$
and $\Pi_{J}:=\openone_{n_{J}}\otimes\openone_{d_{J}}$ are the projectors
on the ${\mathbf{C}}^{n_{J}}\otimes{\mathbf{C}}^{d_{J}}$ sectors
of ${\cal H}.$ In Ref.~\cite{zanardi-dissipation-2014} we have
shown that 
\begin{equation}
{\cal K}_{{\textrm{eff}}}|_{{\mathrm{Ker}}\,{\cal L}_{0}}=-i[{K}_{{\textrm{eff}}},\,\bullet],\qquad{K}_{{\textrm{eff}}}:={\cal P}_{0}({K})\in{\cal A}^{\prime}.\label{K_eff-interaction-alg}
\end{equation}
The effective Hamiltonian ${\cal P}_{0}({K})$ clearly commutes with the
whole unitary group of the interaction algebra. In this sense $K_{\mathrm{eff}}$
is a dissipation-projection symmetrized \cite{symm} version of $K.$
As a consequence, its action is trivial on the ``noise-full''
${\mathbf{C}}^{d_{J}}$ factors in (\ref{state-space-decomp}). In
other terms dissipation can also be regarded as a resource to the
end of {\em{dynamical decoupling}} \cite{symm,symm1,dyn-dec,dyn-dec1}.
\begin{figure}
\noindent \begin{centering}
\includegraphics{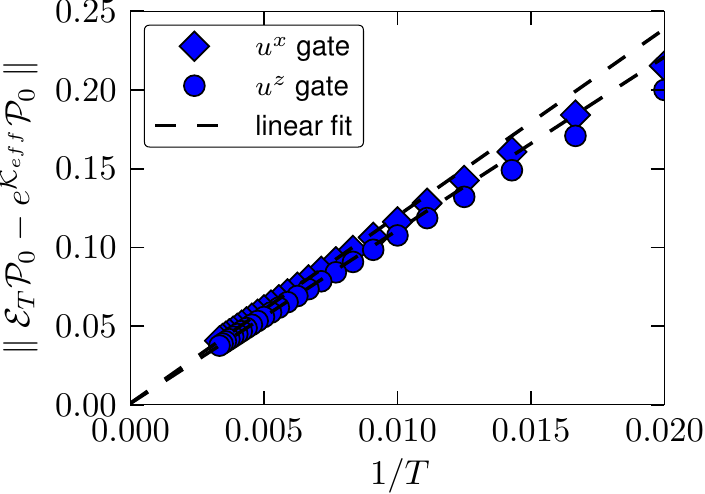} 
\par\end{centering}

\protect\protect\protect\caption{Robustness against dissipative errors. Distance from the exact evolution
and the effective one as a function of $1/T.$ The unperturbed Lindbladian
is of the form $\mathcal{L}_{0}=\sum_{\alpha}\gamma_{\alpha}\mathcal{L}^{\alpha}$,
where $\mathcal{L}^{\alpha}$ is generated by a collective Lindblad
operator $S^{\alpha}=\sum_{j=1}^{N}\sigma_{j}^{\alpha}$ with $N=4$.
The control Hamiltonians are $\mathcal{K}^{\alpha}=-i\left[H^{\alpha},\bullet\right]$
with $H^{x}=(3/2)\left(\sigma_{1}^{z}\sigma_{2}^{z}+\sigma_{2}^{z}\sigma_{3}^{z}\right)+\openone$
and $H^{z}=-(\sqrt{3}/2)\left(\sigma_{1}^{z}\sigma_{2}^{z}-\sigma_{2}^{z}\sigma_{3}^{z}\right)+\sigma_{1}^{z}$.
The effective dynamics generate the unitary gates $u^{\alpha}=\exp{(-i\theta\sigma^{\alpha})}$
with an arbitrary angle $\vartheta$ up to an error $O(T^{-1})$ (see
also \cite{zanardi-dissipation-2014}). The same dynamics is obtained
(up to an error $O(T^{-1})$ with possibly a different prefactor)
with an error on $\mathcal{L}_{0}$ which replaces $S^{\alpha}\to S^{\alpha}+T^{-1}X^{\alpha}$.
For the numerical simulation we used $X^{\alpha}=g\boldsymbol{\sigma}_{1}\cdot\boldsymbol{\sigma}_{2}S^{z}$.
The plot is obtained fixing $\gamma^{\alpha}=g=\vartheta=1$. The
norm used is the maximum singular value of the maps realized as matrices
over ${\cal H}^{\otimes\,2}.$ The linear fit is obtained using the
four most significant points. \label{fig:dissipative-robust}}
\end{figure}


\section{Robustness}

One of the main motivations behind the type of dissipation-assisted
manipulations we are considering, is that it features a significant
degree of built-in resilience. This means that dissipation, besides
providing assistance for QIP, may provide {\em{protection}}.
This stems from the simple observation that the projection theorem
(\ref{projection-th}) clearly indicates that any extra term ${\cal V}$ in the Liouvillian,
either Hamiltonian or dissipative, such that $\|{\cal V}\|=O(1/T)=$ and 
\begin{equation}
{\cal P}_{0}\,{\cal V}\,{\cal P}_{0}=0,\label{robust-eq}
\end{equation}
will not contribute to the effective dynamics (\ref{projection-th}).
For instance in the context of {\em{Example 0}}   any pair of
Hamiltonians $K_{1}$ and $K_{2}$ such that $\mathrm{Tr}_{B}[\rho_{B}(K_{1}-K_{2})]=\lambda\openone_{S},\,(\lambda\in{\mathbf{R}})$
generate the same projected dynamics. 

\subsection{Hamiltonian perturbations}

In the Interaction Algebra case associated with the Liouvillian in Eq.~(\ref{Lindblad-eq}),
one can prove the following result which is  reminiscent of the correctability condition in operator error correction \cite{OPC} (see e.g., Eq. (4) therein).
\vskip 0.2truecm
{\bf{Proposition 2.--}} 
Eq.~ (\ref{robust-eq}) is satisfied by an Hamiltonian perturbation
$V$ iff  \begin{equation}
{\cal P}_{0}({V})\in{\cal A}\cap{\cal A}^{\prime}=:{\cal Z}({\cal A})\label{KL}.
\end{equation}
The solution space of the Hamiltonian
robustness Eq.~(\ref{KL}) is a linear subspace of the full operator
algebra ${\mathrm{L}}({\cal H})$ with codimension $\sum_{J}(n_{J}^{2}-1).$
This subspace, in particular, contains  the kernel of ${\cal P}_{0}$
and the interaction algebra ${\cal A}.$ 

{\em{Proof.--}} 
From Eq.~(\ref{K_eff-interaction-alg}) we see that the condition
(\ref{robust-eq}), for ${\cal V}=-i[V,\bullet],$ means $[{\cal P}_{0}(V),\,\rho]=0,\,\forall\rho\in{\cal A}^{\prime},$
namely the projected dynamics does not change by perturbing $K$ with
any term $V$ such that $\|V\|=O(1/T)$ and ${\cal P}_{0}(V)\in{\cal A}^{\prime\prime}={\cal A}.$
Since, by construction ${\cal P}_{0}(V)\in{\cal A}^{\prime}$ as well
one finds that (\ref{robust-eq}) is satisfied by an Hamiltonian perturbation
$V$ iff Eq.~(\ref{KL}) is satisfied.
Moreover, if $V\in{\cal Z}({\cal A})\Rightarrow\,{\cal P}_{0}(V)=V$
implies that the solution space of Eq.~(\ref{KL}) is the linear space ${\mathrm{Ker}}\,{\cal P}_{0}+{\cal Z}({\cal A}).$
More concretely, the Hamiltonian perturbations $V$ fulfilling the
robustness condition Eq.~ (\ref{KL}) have the form 
\begin{equation}
V=V^{\textrm{off}}+\sum_{J\beta}X_{J}^{\beta}\otimes Y_{J}^{\beta}+\sum_{J}\lambda_{J}\openone_{n_{J}}\otimes\openone_{d_{J}}\label{KerP_0}
\end{equation}
where ${\mathrm{Tr}}(Y_{J}^{\beta})=0,\,(\forall J,\beta),$ and $V^{\textrm{off}}$is
off-diagonal in the decomposition (\ref{alg-decomp}). The first two
terms in the (\ref{KerP_0}) represent ${\mathrm{Ker}}\,{\cal P}_{0}$
whose dimension is then 
$\sum_{J\neq J^{\prime}} (n_{J}d_{J})(n_{J}^{\prime}d_{J}^{\prime})+\sum_{J}n_{J}^{2}(d_{j}^{2}-1)=(\sum_{J}n_{J}d_{J})^{2}-\sum_{J}n_{J}^{2}
= {\mathrm{dim}}\,{\mathrm{L}}({\cal H})-{\mathrm{dim}}\,{\cal A}^{\prime}.$
The third term in (\ref{KerP_0}) represent the center ${\cal Z}({\cal A})$
of the interaction algebra whose dimension is $\sum_{J}1.$ Overall
we see that the solution space of (\ref{KL}) has dimension ${\mathrm{dim}}\,{\mathrm{L}}({\cal H})-\sum_{J}n_{J}^{2}+\sum_{J}1={\mathrm{dim}}\,{\mathrm{L}}({\cal H})-\sum_{J}(n_{J}^{2}-1)$
i.e., it has codimension $\sum_{J}(n_{J}^{2}-1).$ $\hfill\Box$
\vskip 0.2truecm
For example, in the collective decoherence case the interaction algebra $\cal A$ is the algebra of permutation-invariant
operators acting on the $N$-qubit space \cite{DFS,DFS1}. Since 
\begin{eqnarray}
{\cal P}_{0}(\sigma_{j}^{\alpha})&=&\int_{SU(2)} dU U^{\otimes\,N} \sigma_{j}^{\alpha}U^{\dagger\otimes\,N}=
\int_{SU(2)} dU U^{} \sigma_{j}^{\alpha}U^{\dagger}\nonumber \\
&=&{\mathrm{Tr}}( \sigma_{j}^{\alpha})\,\openone=
0\qquad (\alpha=x,y,z;\, j=1,\ldots,N),\nonumber
\end{eqnarray}
it follows that all symmetry-breaking $V$'s of the form $V=\sum_{j=1.\alpha=x,y,z}^{N}\delta_{\alpha}^{j}\sigma_{j}^{\alpha}$ with $|\delta_\alpha^j|=O(1/T) $
can be tolerated. 


\subsection{Dissipative perturbations}

Besides unitary perturbations $K\to K+V$ in practical applications one has also to consider dissipative
ones ${\cal L}_{0}\to{\cal L}_{0}+{\cal L}_{1}$ where ${\cal L}_{1}$
denotes a dissipative Liouvillian with $\|{\cal L}_{1}\|=O(1/T).$
{{It is important to stress that the resilience of the projected
dynamics extends to non-unitary perturbations e.g., extra noise sources.}}

To begin with, we observe that in the unit-preserving case 
all Lindbladian perturbations ${\cal L}_{1}$ of the form of Eq.~(\ref{Lindblad-eq}) 
whose Lindblad operators are in the interaction algebra ${\cal A}$
[see Eq.~(\ref{algs}], satisfy ${\cal P}_{0}{\cal L}_{1}{\cal P}_{0}=0$.
Let us then consider perturbations that take the Lindblad operators outside
of ${\cal A}$. More precisely, we consider Eq.~(\ref{Lindblad-eq})
with Lindblad operators given by collective spin operators $S^{\mu}=\sum_{j=1}^{N}\sigma_{j}^{\mu}\,(\mu=x,y,z)$
and then we perturb them by permutational symmetry breaking terms
${S}^{\mu}\mapsto S^{\mu}+T^{-1}X^{\mu}$ where $\|X^{\mu}\|=O(1).$
This leads to a perturbed Liouvillian $\tilde{{\cal L}}_{0}={\cal L}_{0}+T^{-1}{\cal L}_{1}+T^{-2}{\cal L}_{2}.$
Where 
\begin{equation}
{\cal L}_{1}(\rho)=\sum_{\mu}\left(X^{\mu}\rho S^{\mu}+S^{\mu}\rho X^{\mu}-\frac{1}{2}\{\,\{X^{\mu},\, S^{\mu}\},\,\rho\}\right)\label{linear-correction}
\end{equation}
and ${\cal L}_{2}$ is a quadratic expression in the $X^{\mu}$'s.
\vskip 0.2truecm
{\bf{Proposition 3.--}} If ${\cal L}_{1}$ is given by (\ref{linear-correction})
then ${\cal P}_{0}\,{\cal L}_{1}\,{\cal P}_{0}=0.$

{\em{Proof.--}} To see that is enough to notice that $\rho\in{\cal A}^{\prime}\Rightarrow\rho=\oplus_{J}\rho_{J}\otimes\openone_{d_{J}},$
and $S^{\mu}=\oplus_{J}\openone_{n_{J}}\otimes S_{J}^{\mu}.$ Since ${\cal P}_0$ annihilates any off-diagonal contribution in (\ref{linear-correction})  
 one can assume a $J$ block-diagonal structure
for the perturbation: $X^{\mu}=\oplus_{J}X_{J}^{\mu}\otimes Y_{J}^{\mu}.$
Therefore by considering, for example, the term ${\cal P}_{0}(X^{\mu}\rho S^{\mu})={\cal P}_{0}(X^{\mu}S^{\mu}\rho)={\cal P}_{0}(X^{\alpha}S^{\mu})\rho$
one obtains 
\begin{eqnarray}
{\cal P}_{0}(X^{\mu}S^{\mu}) & = & {\cal P}_{0}\left(\oplus X_{J}^{\mu}\otimes Y_{J}S_{J}^{\mu}\right)=\oplus_{J}{\rm {Tr}}_{d_{J}}(Y_{J}S_{J}^{\mu})\, X_{J}^{\alpha}\otimes\openone/d_{J}\nonumber \\
 & = & \oplus_{J}{\rm {Tr}}_{d_{J}}(S_{J}^{\mu}Y_{J})\, X_{J}^{\mu}\otimes\openone/d_{J}={\cal P}_{0}(S^{\mu}X^{\mu}).
\end{eqnarray}
This shows that any term in ${\cal P}_{0}\,{\cal L}_{1}\,{\cal P}_{0}$
arising e.g., from the first term in (\ref{linear-correction}), is
canceled by an identical one arising from the anti-commutator side
as ${\cal P}_{0}(X^{\mu}S^{\mu}\rho+S^{\mu}X^{\mu}\rho)=2{\cal P}_{0}(S^{\mu}X^{\mu})\rho.$
Notice that in particular for $X^{\mu}\in{\cal A}^{\prime}$ one has
the stronger property ${\cal L}_{1}{\cal P}_{0}=0.$ $\hfill\Box$

In Fig.~\ref{fig:dissipative-robust} we report a numerical simulation
for the four qubits system discussed in the former section. The simulation
confirms that for small $1/T$ the Liouvillians ${\cal L}_{0}$ and
$\tilde{{\cal L}}_{0}$ generate the same projected dynamics \cite{zanardi-dissipation-2014}.
In words: one can exploit (symmetric) noise to wash out other noise.



\section{Emerging unitarity}

\begin{figure}
\noindent \begin{centering}
\includegraphics[height=3.5cm]{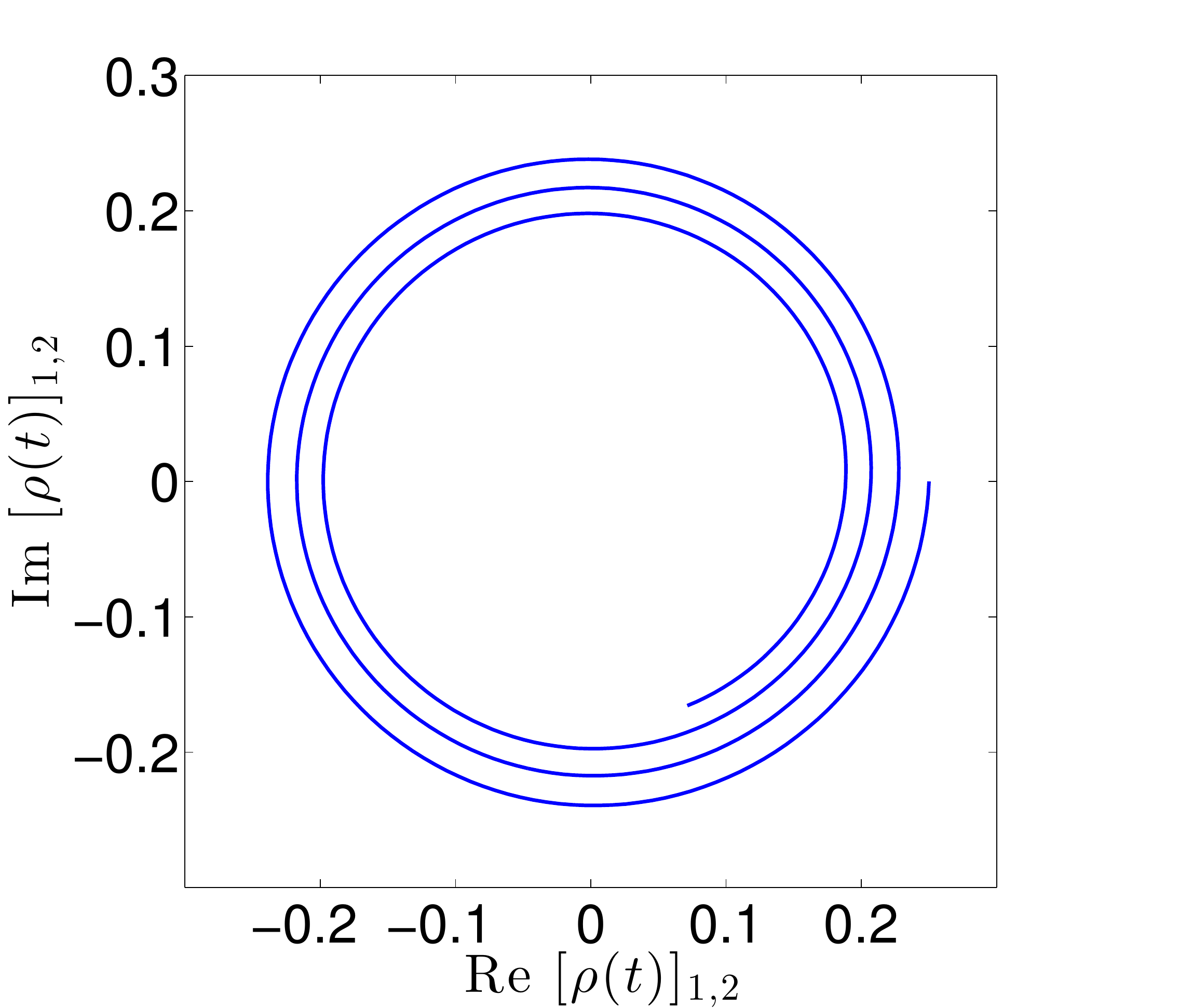}\includegraphics[height=3.4cm]{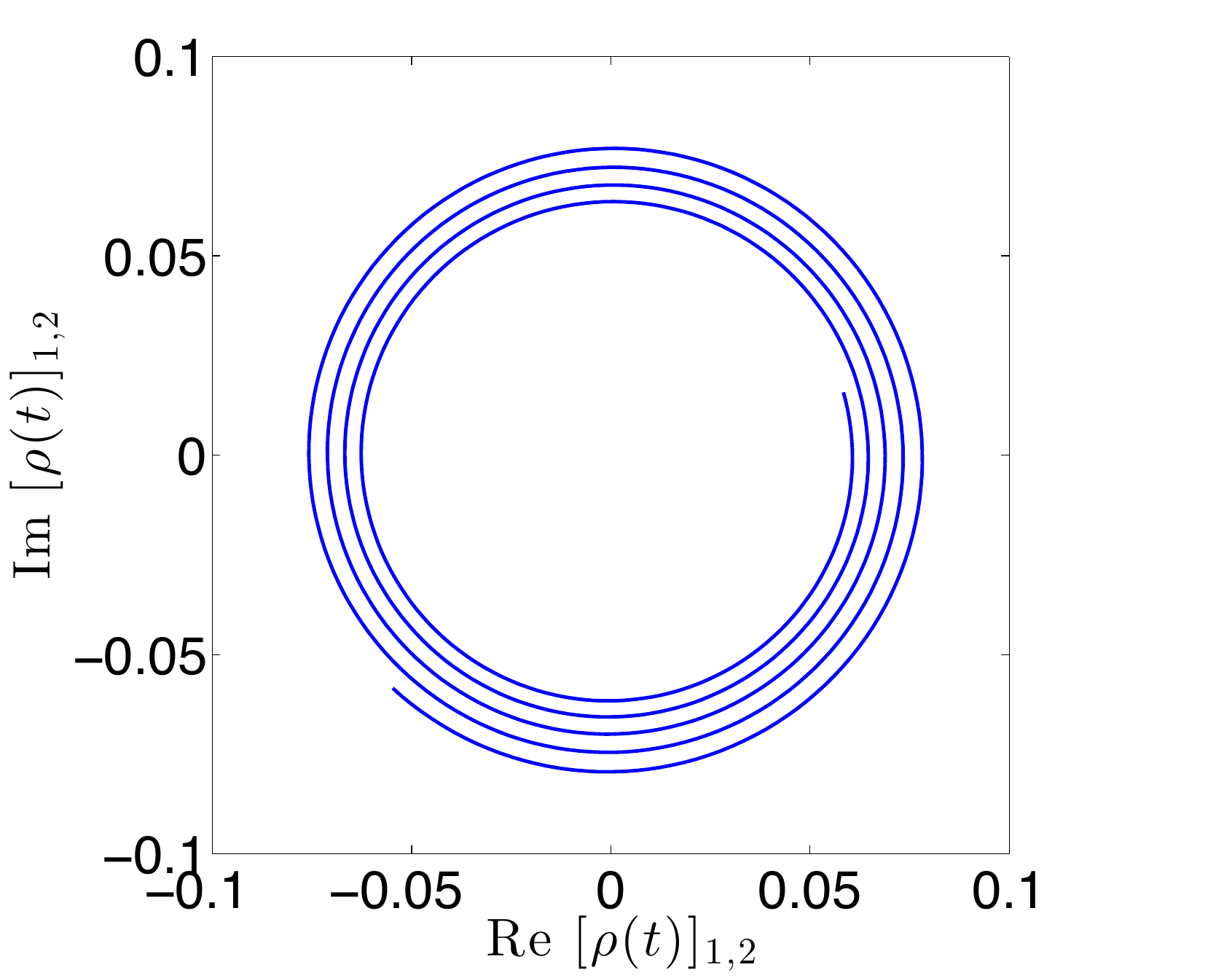} 
\par\end{centering}

\protect\caption{Unitarity from dissipation: a unitary dynamics is generated out of
a purely dissipative one. To highlight the unitary character we use
the method described in \cite{Q-BIO}. For unitary dynamics one has
$\rho(t)=e^{-itH_{\mathrm{eff}}}\rho_{0}e^{itH_{\mathrm{eff}}}=\sum_{n,m}e^{-it(E_{n}-E_{m})}\left[\rho_{0}\right]_{n,m}|n\rangle\langle m|$,
with $H_{\mathrm{eff}}=\sum_{n}E_{n}|n\rangle\langle n|$. In the
figure we plot the real and imaginary part of $\langle1|\rho(t)|2\rangle$.
Left panel: two qubit example generating the effective Hamiltonian
$H_{eff}=\sigma^{y}\otimes\openone$. The time evolution window is
$t\in[T/100,10T]$ and $T=100$ arbitrary units (AU). Right panel:
four qubit example with perturbed collective noise. The effective
Hamiltonian is given in Eq.~\ref{eq:Heff4}. The time evolution window
is $[T/2,4T]$ with $T=100$ AU.\label{fig:Unitarity-from-dissipation.}}
\end{figure}

A unitary dynamics gives rise to a non-unitary one as soon as some
unobserved degrees of freedom are traced out. This is an ubiquitous
situation in physics. The converse process, to obtain a unitary evolution
from an underlying dissipative one, appears a much more difficult
task. Here we show how this phenomenon of {\em {emerging unitarity}}
manifests itself in the context we have discussed in this paper.

We begin by slightly generalizing our set-up i.e., going beyond the
unital case where the kernel of ${\cal L}_{0}$ is the commutant of
the interaction algebra. According to Ref ~\cite{baum} any generator
of the Lindblad type (\ref{Lindblad-eq}) is such that its SSS is
given by states $\rho$ of the form $\rho=\sum_{J}^{\oplus}\lambda_{J}\,\omega_{J}\otimes\rho_{0,J}$
where the state space block structure is still of the form (\ref{state-space-decomp}).
Here the  $\omega_{J}$ are arbitrary states in ${\mathbf{C}}^{n_{J}},$
the $\rho_{0,J}$'s are  uniquely defined states in ${\mathbf{C}}^{n_{J}}$
and the $\lambda_{J}$'s are non-negative scalars. In the unital case
we mostly considered so far $\rho_{0,J}=\openone_{d_{J}}/d_{J}.$
The robustness calculation of the the former section can be now generalized
to this non-unital Liouvillians case. As before we perturb the (not
necessarily hermitean) Lindblad operators $L_{\alpha}\mapsto L_{\alpha}+\delta L_{\alpha}$
and consider as perturbation the first order variation of ${\cal L}_{0}.$
\begin{equation}
{\cal L}_{1}(\rho)=\sum_{\alpha}(\delta L_{\alpha}\rho L_{\alpha}^{\dagger}-\frac{1}{2}(\delta L_{\alpha}^{\dagger}L_{\alpha}+L_{\alpha}^{\dagger}\delta L_{\alpha})\,\rho+{\mathrm{h.c.}})\label{L_1}
\end{equation}
If $\|\delta L_\alpha\|=O(1/T)$  then Eq.~(\ref{projection-th}) holds with $\tilde{{\cal K}}$ replaced by ${\cal L}_1.$
\vskip 0.2truecm  
{\bf{Proposition 4.--}} Let us add, to a Liouvillian generator of
the type (\ref{Lindblad-eq}), a perturbation of the form (\ref{L_1}).
Then, for $\rho$ in the SSS, one has that ${\cal P}_{0}{\cal L}_{1}{\cal P}_{0}(\rho)=-i[A,\,\rho],$
where 
\begin{equation}
A={\mathrm{Im}}\,\sum_{\alpha}\sum_{J}^{\oplus}{\mathrm{Tr}}_{d_{J}}\left(\delta L_{\alpha}^{\dagger}L_{\alpha}\,(\openone_{n_{J}}\otimes\rho_{0,J})\right)\otimes\openone_{d_{J}} =A^\dagger\label{eff-Ham-tot}
\end{equation}
in which ${\mathrm{Im}}\, X:=\frac{1}{2i}(X-X^{\dagger}).$
In particular, for the unital case ${\cal L}_{0}(\openone)=0,$ one
can write 
\begin{equation}
A={\mathrm{Im}}\,{\cal P}_{0}\,(\sum_{\alpha}\delta L_{\alpha}^{\dagger}L_{\alpha}).
\label{emergent-unitary}
\end{equation}

{\em{Proof.--}} Let us consider the Lindbladian ${\cal L}_{0}(\rho)=L\rho L^{\dagger}-\frac{1}{2}\{L^{\dagger}L,\,\rho\}.$
Perturbing the Lindblad operators $L\mapsto L+\delta L$ one finds
the variation ${\cal L}_{0}\mapsto{\cal L}_{0}+\delta{\cal L},$ where
\begin{equation}
\delta{\cal L}(\rho)=\delta L\rho L^{\dagger}-\frac{1}{2}(L^{\dagger}\delta L+\delta L^{\dagger}L)\,\rho+{\mathrm{h.c}},\label{deltaL}
\end{equation}
Without loss of generality we can consider $\delta L$ as block-diagonal
in the decomposition (\ref{alg-decomp}) and work on a fixed $J$
sector ${\mathbf{C}}^{n_{J}}\otimes{\mathbf{C}}^{d_{J}}$. In that
sector we write: $\delta L=X\otimes Y,$ $L=\openone\otimes\tilde{L}$
and $\rho=\omega\otimes\rho_{0}$ in the SSS. Here $\rho_{0}$ denotes
the unique steady-state in the ${\mathbf{C}}^{d_{J}}$ factor of the
$J$ block i.e., $\openone_{d_{J}}/d_{J}$ in the unital case. The
first three terms in (\ref{deltaL}) give rise to the following three
contributions respectively 
\[
X\omega\otimes Y\rho_{0}\tilde{L}^{\dagger},\;\;\;-\frac{1}{2}X\omega\otimes\tilde{L}^{\dagger}Y\rho_{0},\;\;\;-\frac{1}{2}X^{\dagger}\omega\otimes Y^{\dagger}\tilde{L}\,\rho_{0}.
\]

Applying ${\cal P}_{0}\colon A\otimes B\mapsto{\mathrm{Tr}}(B)\, A\otimes\rho_{0}$
and adding the h.c., terms one finds 
\begin{align}
&{\cal P}_{0}\delta{\cal L}{\cal P}_{0}(\rho) =(\frac{\alpha}{2}X\omega-\frac{\bar{\alpha}}{2}X^{\dagger}\omega)\otimes\rho_{0}+{\mathrm{h.c.}} =\nonumber \\
 & -i\,[\frac{\bar{\alpha}X^{\dagger}-{\alpha}X}{2i},\,\omega]\otimes\rho_{0},\quad\alpha:={\mathrm{Tr}}(\tilde{L}^{\dagger}Y\rho_{0}).\label{eff-Ham}
\end{align}
On the other hand $\delta L^{\dagger}L=(X^{\dagger}\otimes Y^{\dagger})(\openone\otimes\tilde{L})=X^{\dagger}\otimes Y^{\dagger}\tilde{L}$
from which we see that (\ref{eff-Ham}) can be written as $-i[\tilde{A},\,\omega\otimes\rho_{0}]$
where $\tilde{A}={\mathrm{Im}}\,{\mathrm{Tr}}_{d_{J}}\left(\delta L^{\dagger}L\,(\openone\otimes\rho_{0})\right)\otimes\openone_{d_{J}}.$
Here the index $d_{J}$ denotes the second factor i.e., ${\mathbf{C}}^{d_{J}}$ in the bipartition
of the given $J$ block. Let us now consider a Liouvillian ${\cal L}_{0}$
with more than one Lindblad operator $\{L_{\alpha}\}_{\alpha}$. Putting
together all the different $J$-blocks, we obtain ${\cal P}_{0}{\cal L}_{1}{\cal P}_{0}(\rho)=-i[A,\,\rho]$
where $A$ is given by Eq.~(\ref{eff-Ham-tot}) with $\rho=\sum_{J}^{\oplus}\omega_{J}\otimes\rho_{0,J},\,({\cal P}_{0}(\rho)=\rho).$
In the unital case one has $\rho_{0,,J}={\openone_{d_{J}}}/{d_{J}}$
and Eq.~(\ref{emergent-unitary}) follows from (\ref{eff-Ham-tot})
and (\ref{alg-projection}).$\hfill\Box$ \vskip 0.2truecm {\bf
{Corollary}} i) If all the $L_{\alpha}$'s and perturbations $\delta L_{\alpha}$'s
are Hermitean $A=0;$ ii) If $\delta L_{\alpha}\in{\cal A}^{\prime}(\Rightarrow(X_{\alpha})_{J}\sim\openone_{n_{J}})$
then $A_{J}\sim\openone_{n_{J}}\Rightarrow{\cal P}_{0}\delta{\cal L}{\cal P}_{0}=0.$
\vskip 0.2truecm

While mathematically simple the Proposition 4 is, on physical grounds, a remarkable and surprising result. 
Combined with the dissipation-projection theorem Eq.~(\ref{projection-th}), it indeed implies that a small, generic, Lindblad perturbation 
induces an effective unitary dynamics over the SSS generated by the Hamiltonian (\ref{eff-Ham-tot}). This 
{\em even in absence of any Hamiltonian term in both the unperturbed
and unperturbed Liouvillian}. In principle, by
tailoring the dissipative terms $\delta L_{\alpha}$'s, one can obtain
a desired effective unitary generator $A.$

\subsection{Examples}

\begin{figure}
\noindent \begin{centering}
\includegraphics[height=3.5cm]{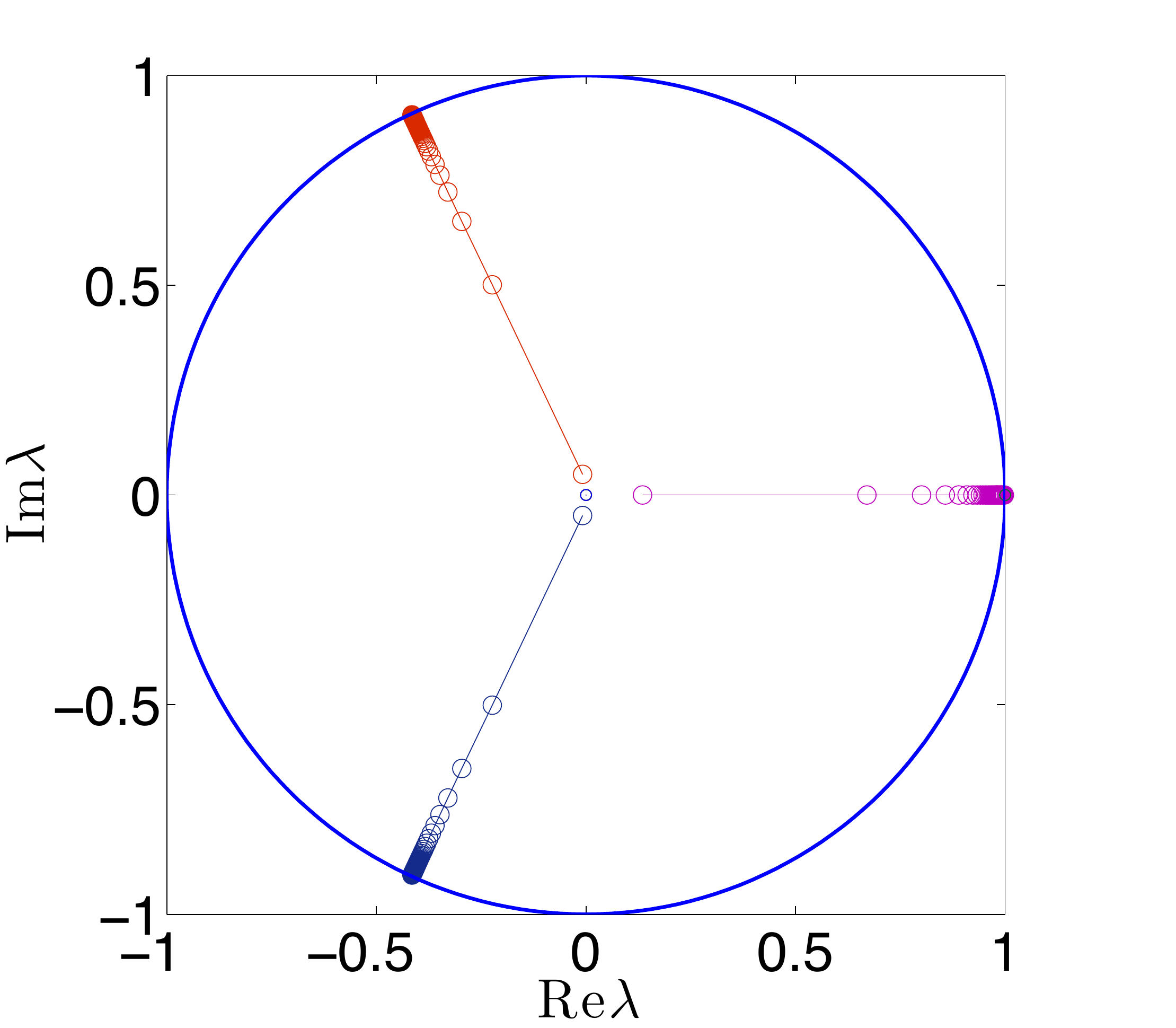}\includegraphics[height=3.5cm]{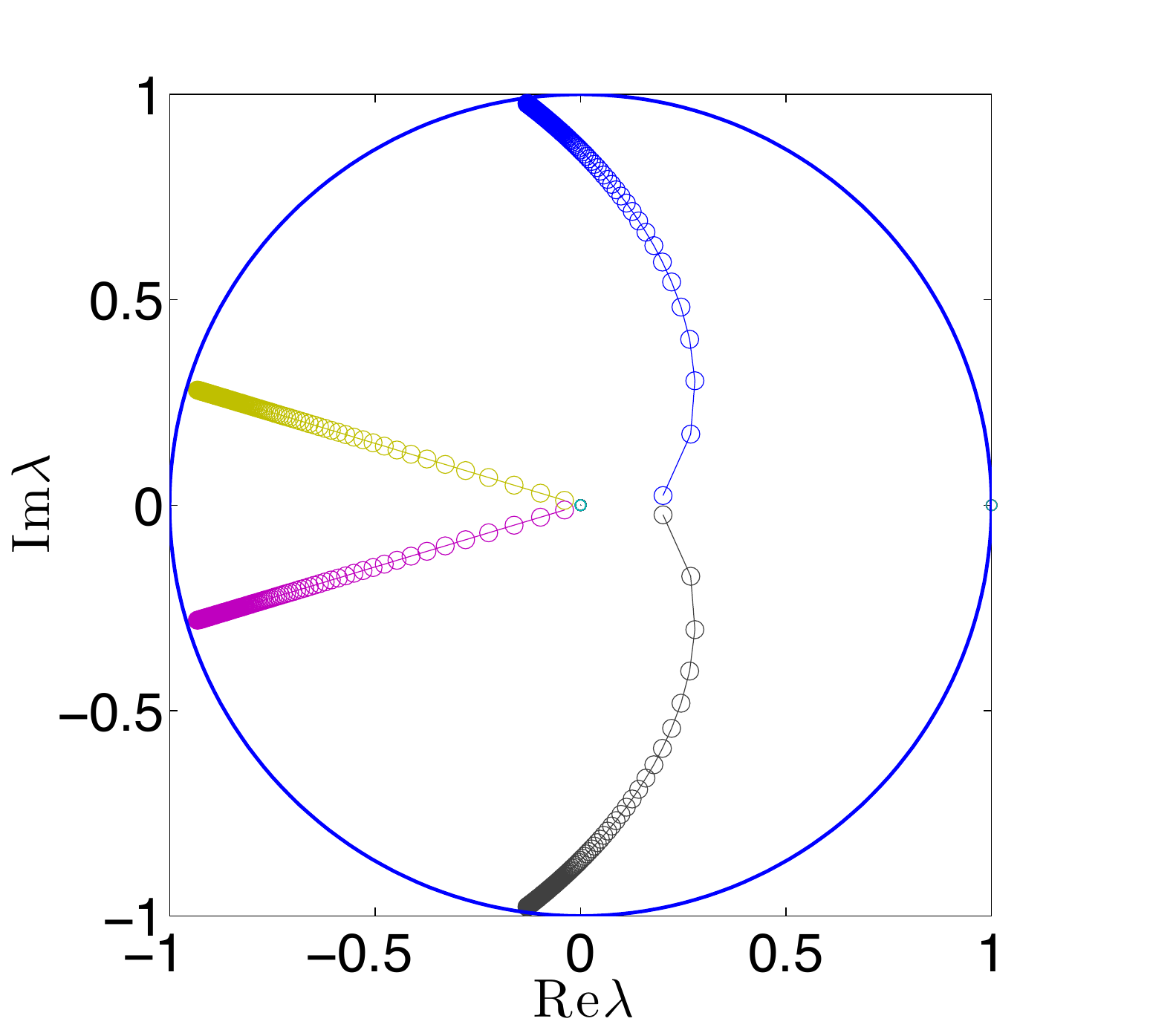} 
\par\end{centering}

\protect\caption{Unitarity from dissipation: a unitary dynamics is generated out of
a purely dissipative one. Here we plot real and imaginary part of
the eigenvalues of the exact map $\mathcal{E}_{T}{\cal P}_{0}=\exp\left[T(\mathcal{L}_{0}+\frac{1}{T}{\cal L}_1 +\frac{1}{T^2}{\cal L}_2)\right]{\cal P}_{0}$ 
for different $T$. The examples are the same as in Fig.~\ref{fig:Unitarity-from-dissipation.},
i.e.~two (four) qubit example on the left (right) panel. Increasing
$T$ the eigenvalues tend to go on the unit circle. Left panel: $T\in[1,1000]$
AU, the eigenvalues approach $e^{\pm2i},1$, consistently with Eq.~(\ref{eq:2qubit_ham}).
Right panel: $T\in[5,600]$ AU. The eigenvalues of the effective Hamiltonians (\ref{eq:Heff4}) are $\pm8,0$. Consequently the eigenvalues
of $\mathcal{E}_{T}P_{0}$ approach $e^{\pm16i},e^{\pm8i},1$. \label{fig:Unitarity-from-dissipation.-1}}
\end{figure}

To illustrate this mechanism we first consider a simple two-qubit
example. We set $L=\openone\otimes\sigma^{z}$ and $\delta L=\sigma^{-}\otimes\sigma^{z}.$
In this case ${\cal P}_{0}(X)={\mathrm{Tr}}_{B}\left(X(\openone\otimes|0\rangle\langle0|)\right)\otimes|0\rangle\langle0|+{\mathrm{Tr}}_{B}\left(X(\openone\otimes|1\rangle\langle1|)\right)\otimes|1\rangle\langle1|,$
where ${\mathrm{Tr}}_{B}$ denotes the partial trace over the second
qubit. Using Eqs.~(\ref{eff-Ham}), (\ref{eff-Ham-tot}) with $J=0,1;\, d_{0}=d_{1}=1;\,\rho_{0,0}=|0\rangle\langle0|,\,\rho_{0,1}=|1\rangle\langle1|;\, X=\sigma^{-}$
and $Y=\sigma^{z}$ one finds 
\begin{equation}
A={\mathrm{Im}}\,{\cal P}_{0} (\delta L^{\dagger}L)=\sigma^{y}\otimes|0\rangle\langle0|+\sigma^{y}\otimes|1\rangle\langle1|=\sigma^{y}\otimes\openone. 
 \label{eq:2qubit_ham} 
\end{equation}

A numerical simulation of this dissipation generated gate is shown
in Figs.~\ref{fig:Unitarity-from-dissipation.}, \ref{fig:Unitarity-from-dissipation.-1} left panel. To highlight
the unitary character of the dynamics we use the fact that matrix
elements of the density matrix $\rho(t)$ evolve as phases in the
Hamiltonian eigenbasis and therefore result in circles on the complex
plane [Fig.~\ref{fig:Unitarity-from-dissipation.}]. Similarly as $\mathcal{E}_T$ approaches a unitary evolution within the SSS, the eigenvalues of $\mathcal{E}_T \mathcal{P}_0$ converge to the unit circle increasing $T$ as depicted in Fig.~\ref{fig:Unitarity-from-dissipation.-1}.

Let us now consider a four-qubit system subject to general collective
decoherence \cite{DFS,DFS1}. In this case ${\cal L}_{0}$ has the
form (\ref{Lindblad-eq}) with the Lindblad operators given by collective
spin operators i.e., $L_{\alpha}=\sum_{j=1}^{4}\sigma_{j}^{\alpha},\,(\alpha=x,y,z).$
In this unital case the interaction algebra ${\cal A}$ coincides
with algebra of totally symmetric operators and the commutant ${\cal A}^{\prime}$
is $14-$dimensional \cite{zanardi-dissipation-2014} and generated
by qubit permutation operators in the group ${\cal S}_{4}$. We consider
perturbations of the form $\delta L_{\alpha}=U\,{L}_{\alpha}=L_{\alpha}\, U$
where $U\in{\cal S}_{4}.$ Then $\sum_{\alpha}\delta L_{\alpha}^{\dagger}L_{\alpha}=4U^{\dagger}\boldsymbol{S}^{2}$, where $\boldsymbol{S}$ is the total spin operator.  
We also have $\mathcal{P}_{0}\left(U^{\dagger}\boldsymbol{S}^{2}\right)=U^{\dagger}\boldsymbol{S}^{2}$
and $U^{\dagger}\boldsymbol{S}^{2}-U\boldsymbol{S}^{2}=2(U^{\dagger}-U)$.
We further fix $U$  to perform the right-shift
permutation $\left(1,2,3,4\right)\to\left(4,1,2,3\right)$. One obtains
\begin{align}
A & =4\,\mathrm{Im}\,\mathcal{P}_{0}\left(U^{\dagger}\boldsymbol{S}^{2}\right)=8\,\mathrm{Im}(U^{\dagger})\\
 & =\left[(\boldsymbol{\sigma}_{1}+\boldsymbol{\sigma}_{2})\times\boldsymbol{\sigma}_{3}\right]\cdot\boldsymbol{\sigma}_{4}+\left[\boldsymbol{\sigma}_{2}\times(\boldsymbol{\sigma}_{3}+\boldsymbol{\sigma}_{4})\right]\cdot\boldsymbol{\sigma}_{1}\,,\label{eq:Heff4}
\end{align}
where in the last equation we used the fact that $U=S_{2,3}S_{3,4}S_{1,4}$
with $S_{i,j}=\left(\openone+\boldsymbol{\sigma}_{i}\cdot\boldsymbol{\sigma}_{j}\right)/2$
the operator swapping site $i$ with $j$. A numerical simulation confirming this unitary behavior emerging from a dissipative dynamics, 
is shown in Figs.~\ref{fig:Unitarity-from-dissipation.},\ref{fig:Unitarity-from-dissipation.-1} right panel.

\section{Conclusions}

The traditional avenue to quantum information processing (QIP)  primitives, such
as quantum gates, requires the dissipation due to
the environment to be as small a possible compared to the control Hamiltonian. 
A number of powerful techniques have been develop to combat the detrimental effects of dissipation  \cite{QEC}.
However, over the last few years there has been a growing amount of evidence that dissipation may on the contrary provide a {\em{resource}} for QIP  see e.g., \cite{Kraus-prep,kastoryano2011dissipative,barreiro2011open,verstraete2009quantum,dissi-top,gauge}.

In this spirit in ref.~\cite{zanardi-dissipation-2014} we
have shown  how it is possible to generate coherent quantum manipulations also in the opposite
regime in which the dissipation is much stronger than the control
Hamiltonian. The only requirement is essentially that the dissipation
must provide a degenerate set of  steady states (SSS).
The coherent control drives the system away from the SSS but the strong
dissipation effectively projects the dynamics back onto the SSS. As
a consequence a quantum evolution governed by an effective Hamiltonian
coherently unfolds  within the SSS \cite{zanardi-dissipation-2014}. 

In this paper we further investigated the consequences of this approach. The following are the main findings
of this paper. {\em{ i)}} We provided further details on the  rigorous  estimate of the error
between the exact evolution and the effective projected dynamics. {\em{ii)}} Moving  to a suitable rotated frame, we have shown 
that the effective dynamics in  the SSS is of geometric origin i.e., it is the  holonomy
associated with a superoperator-valued connection. {\em{iii)}} The effective dynamics
is protected against a large class of Hamiltonian and dissipative perturbations. 
{\em{iv)}} 
As a corollary of this result, we have shown that certain dissipative perturbations
of purely, dissipative, Lindbladian (i.e.~one for which the eigenvalues
are real negative)  generate an effective  unitary dynamics. 

Dissipative dynamics is easily obtained from a unitary one 
as soon as some degrees of freedom are traced out. On the contrary
the emergent unitarity phenomenon we have discussed 
is a quite surprising  example of a unitary dynamics
obtained from a purely dissipative one. Understanding its fundamental origin
and potential application in QIP is a topic worthwhile
of future investigations.

\begin{acknowledgments}
This work was partially supported by the ARO MURI grant W911NF-11-
1-0268. We thank I. Marvian for useful discussions.
\end{acknowledgments}


\newpage

\appendix

\section{Error estimate\label{sec:Error-estimate}}

Here we are going to estimate the error term appearing in the RHS
of Eq.~(\ref{projection-th}). In this section we use the following
notation for the total Liouvillian: $\mathcal{L}(x)=\mathcal{L}_{0}+x\mathcal{L}_{1}$,
where $\mathcal{L}_{0}$ is the dominant, dissipative, term, $\mathcal{L}_{1}$
is the perturbation (which often will be taken as a unitary generator,
i.e.~$\mathcal{L}_{1}=\mathcal{K}$), and $x$ is a small dimensionless
parameter. The relation between $x$ and the $T$ in the main text
is $T^{-1}=x\tau_{0}^{-1}$ where $\tau_{0}$ is some time-scale which will become more explicit below.
We are going to show that the LHS of Eq.~(\ref{projection-th}) is
analytic in $x$ around zero starting with a linear term and we are
going to estimate its coefficient. Let us assume that $\mathrm{Ker}(\mathcal{L}_{0})$
is $d$-dimensional, i.e.~$d$ eigenvalues of $\mathcal{L}_{0}$
are zero. If we turn on the perturbation $x\mathcal{L}_{1}$, some
of these eigenvalues will move a little bit. The collection of all
these $d$ eigenvalues forms the so called $\lambda$-group \cite{kato}
and identifies an invariant subspace of $\mathcal{L}(x)$. The projection
$\mathcal{P}(x)$ onto such subspace turns out to be an analytic function
of $x$ \cite{kato}. As shown in \cite{kato}, the restriction of
$\mathcal{L}(x)$ to the $\lambda$-group, $\mathcal{R}(x):=\mathcal{P}(x)\mathcal{L}(x)=\mathcal{L}(x)\mathcal{P}(x)=\mathcal{P}(x)\mathcal{L}(x)\mathcal{P}(x)$,
is also an analytic function of $x$. Then one has the following expansions
\begin{eqnarray}
\mathcal{P}(x) & = & \mathcal{P}_{0}+x\mathcal{P}_{1}+O(x^{2})\\
\mathcal{R}(x) & = & x\mathcal{R}_{1}+x^{2}\mathcal{R}_{2}+x^{3}\mathcal{R}_{3}+O(x^{4}).
\end{eqnarray}
 The Liouvillian character of $\mathcal{L}_{0}$ (i.e.~that fact
that $\mathcal{P}_{0}=\lim_{t\to\infty}e^{t\mathcal{L}}$) assures
that zero is a semisimple eigenvalue of $\mathcal{L}_{0}$, i.e.~there
is no Jordan block associated with the zero eigenvalue of $\mathcal{L}_{0}$.
Whereas this is not strictly required it does simplify the following
formulae. In the semisimple case one has for instance \cite{kato}:
\begin{align}
\mathcal{P}_{1} & =  -(\mathcal{P}_{0}\mathcal{L}_{1}\mathcal{S}+\mathcal{S}\mathcal{L}_{1}\mathcal{P}_{0})\\
\mathcal{R}_{1} & =  \mathcal{P}_{0}\mathcal{L}_{1}\mathcal{P}_{0}\\
\mathcal{R}_{2} & =  -\left(\mathcal{P}_{0}\mathcal{L}_{1}\mathcal{P}_{0}\mathcal{L}_{1}\mathcal{S}+\mathcal{P}_{0}\mathcal{L}_{1}\mathcal{S}\mathcal{L}_{1}\mathcal{P}_{0}+\mathcal{S}\mathcal{L}_{1}\mathcal{P}_{0}\mathcal{L}_{1}\mathcal{P}_{0}\right).
\end{align}
In the above formulae, $\mathcal{S}$ is the projected resolvent of
$\mathcal{L}_{0}$ at zero satisfying $\mathcal{L}_{0}\mathcal{S}=\mathcal{S}\mathcal{L}_{0}=\openone-\mathcal{P}_{0}$
and is given by
\begin{equation}
\mathcal{S}=-\sum_{k\neq0}\left[(-\lambda_{k})^{-1}\mathcal{P}^{(k)}+\sum_{n=1}^{m_{k}-1}(-\lambda_{k})^{-n-1}\left(\mathcal{D}^{(k)}\right)^{n}\right],
\end{equation}
assuming that $\mathcal{L}_{0}$ has Jordan decomposition 
\begin{equation}
\mathcal{L}_{0}=\sum_{k}\left(\lambda_{k}\mathcal{P}^{(k)}+\mathcal{D}^{(k)}\right)
\end{equation}
with $\lambda_{k}$ eigenvalues with (algebraic) multiplicity $m_{k}$,
projectors $\mathcal{P}^{(k)}$ and nilpotent blocks $\mathcal{D}^{(k)}$
(note that $\lambda_{0}=0$ and $\mathcal{P}^{(0)}\equiv\mathcal{P}_{0}$).
Note that all the $\mathcal{L}_{j},\,\mathcal{R}_{j}'s$ have the
dimension of Hz, the $\mathcal{P}_{j}$'s are dimensionless and $\mathcal{S}$
has units of time. In particular $tx\mathcal{R}_{1}$ is precisely
$\mathcal{L}_{\mathrm{eff}}$ in our applications. We denote it $\mathcal{R}_{eff}:=tx\mathcal{R}_{1}$
here for notational consistency. 

Clearly $\mathcal{P}(x)$ commutes with $\mathcal{L}(x)$ and so one
has the identity
\begin{equation}
e^{t\mathcal{L}(x)}\mathcal{P}(x)=e^{t\mathcal{R}(x)}\mathcal{P}(x).
\end{equation}

We now choose times $t$ such that $tx$ is bounded by a given finite
time in some unit, i.e.~$tx=O(1)\tau_{0}$. Note that $\left\Vert e^{t\mathcal{L}(x)}\right\Vert \le1$
because the evolution maps states to states (i.e.~because of positivity).
Instead from $\left\Vert e^{t\mathcal{R}(x)}\right\Vert \le\exp\left\Vert t\mathcal{R}(x)\right\Vert $
we get $\left\Vert e^{t\mathcal{R}(x)}\right\Vert \le O(1)$. Hence
we obtain 
\begin{equation}
\left(e^{t\mathcal{L}(x)}-e^{t\mathcal{R}(x)}\right)\mathcal{P}_{0}=-\left(e^{t\mathcal{L}(x)}-e^{t\mathcal{R}(x)}\right)x\mathcal{P}_{1}+O(x^{2}).
\end{equation}

Now define $\Delta=:e^{tx\mathcal{R}_{1}}-e^{t\mathcal{R}(x)}$. Clearly
$\Delta=O(x)$ (we will later determine the coefficient), so we finally
get
\begin{equation}
\left(e^{t\mathcal{L}(x)}-e^{tx\mathcal{R}_{1}}\right)\mathcal{P}_{0}=+\Delta\mathcal{P}_{0}-\left(e^{t\mathcal{L}(x)}-e^{tx\mathcal{R}_{1}}\right)x\mathcal{P}_{1}+O\left(x^{2}\right).\label{eq:bound_start}
\end{equation}
Using Dyson expansion one can easily estimate $\Delta$:
\[
\Delta=tx^{2}\int_{0}^{1}e^{(1-s)\mathcal{R}_{eff}}\mathcal{R}_{2}e^{s\mathcal{R}_{eff}}ds+O\left(x^{2}\right).
\]

Quite naturally we use a submultiplicative and automorfism-invariant
norm for superoperators. We now take the norm of Eq.~(\ref{eq:bound_start}),
use triangle inequality and bound all the resulting terms. Defining
$C=\sup_{s\in\left[0,1\right]}\left\Vert e^{s\mathcal{R}_{\mathrm{eff}}}\right\Vert $,
we get $\left\Vert \Delta\right\Vert \le tx^{2}C^{2}\left\Vert \mathcal{R}_{2}\right\Vert +O(x^{2})$,
$\left\Vert \mathcal{R}_{2}\right\Vert \le3\left\Vert \mathcal{P}_{0}\right\Vert ^{2}\left\Vert \mathcal{L}_{1}\right\Vert ^{2}\left\Vert S\right\Vert $,
and $\left\Vert \mathcal{P}_{1}\right\Vert \le2\left\Vert \mathcal{P}_{0}\right\Vert \left\Vert \mathcal{L}_{1}\right\Vert \left\Vert S\right\Vert $.
Putting things together we finally obtain
\begin{multline}
\left\Vert \left(e^{t\mathcal{L}(x)}-e^{\mathcal{R}_{\mathrm{eff}}}\right)\mathcal{P}_{0}\right\Vert \le x\left\Vert \mathcal{S}\right\Vert \left\Vert \mathcal{L}_{1}\right\Vert \left\Vert \mathcal{P}_{0}\right\Vert \\
\times\left(3txC^{2}\left\Vert \mathcal{P}_{0}\right\Vert ^{2}\left\Vert \mathcal{L}_{1}\right\Vert +4\right)+O(x^{2}).\label{eq:bound_final}
\end{multline}

In order to make more apparent the connection with physical constants,
we define dimensionless (tilded) operators via $\mathcal{L}=\gamma_{0}\tilde{\mathcal{L}}_{0}+\gamma_{1}\tilde{\mathcal{L}}_{1}$
such that $\gamma_{0}^{-1}=\tau_{R}$ is the (short) relaxation time
of the unperturbed dissipation and $\gamma_{1}^{-1}=T$ is the timescale
of the control term. Measuring times in units of $\tau_{R}$ the evolution
becomes $\exp\left[(t/\tau_{R})\left(\tilde{\mathcal{L}}_{0}+x\tilde{\mathcal{L}}_{1}\right)\right]$
and we see that $x=\gamma_{1}/\gamma_{0}=\tau_{R}/T$ is the small
parameter. The requirement that the effective generator $\tilde{\mathcal{R}}_{\mathrm{eff}}=t\gamma_{1}\tilde{\mathcal{P}}_{0}\tilde{\mathcal{L}}_{1}\tilde{\mathcal{P}}_{0}$
is finite and non-zero implies $t\gamma_{0}x=t\gamma_{1}=O(1)$. This means that
the waiting time is given by $t=O(T)$. The bound Eq.~(\ref{eq:bound_final})
then translates into (all the tilded operators are dimensionless)
\begin{multline}
\left\Vert \left(e^{t\mathcal{L}}-e^{\tilde{\mathcal{R}}_{\mathrm{eff}}}\right)\mathcal{P}_{0}\right\Vert \le\frac{\tau_{R}}{T}\left\Vert \tilde{\mathcal{S}}\right\Vert \left\Vert \tilde{\mathcal{L}}_{1}\right\Vert \left\Vert \tilde{\mathcal{P}}_{0}\right\Vert \\
\times\left(3\frac{t}{T}C^{2}\left\Vert \tilde{\mathcal{P}}_{0}\right\Vert ^{2}\left\Vert \tilde{\mathcal{L}}_{1}\right\Vert +4\right)+O(x^{2}).\label{eq:bound_final-1}
\end{multline}
Equation above can often be further simplified. For example, if $\mathcal{L}_{0}$
generate a positive map $\left\Vert \tilde{\mathcal{P}}_{0}\right\Vert =1$,
whereas if $\mathcal{R}_{\mathrm{eff}}$ generates a unitary one has
$C=1$.

\end{document}